\documentclass[prd,twocolumn,aps,amsmath,nofootinbib,superscriptaddress]
{revtex4}

\usepackage{graphicx}
\usepackage{bm}
\usepackage{epsfig}

\def\ltap{\;\centeron{\raise.35ex\hbox{$<$}}{\lower.65ex\hbox{$\sim$}}\;}
\def\gtap{\;\centeron{\raise.35ex\hbox{$>$}}{\lower.65ex\hbox{$\sim$}}\;}

\newcommand{\nn}{\nonumber} 
\newcommand{\bea}{\begin{eqnarray}}
\newcommand{\eea}{\end{eqnarray}}

\begin{document}


\preprint{\vbox{\hbox{68-2532} \hbox{hep-ph/0412007} }}

\title{Fluctuating annihilation cross sections and the generation of 
density perturbations}

\author{Christian W.~Bauer}
\affiliation{California Institute of Technology, Pasadena, CA 91125}
\author{Michael L.~Graesser}
\affiliation{California Institute of Technology, Pasadena, CA 91125}
\author{Michael P.~Salem}
\affiliation{California Institute of Technology, Pasadena, CA 91125}


\begin{abstract}
Fluctuations in the mass and decay rate of a heavy particle which for some 
period dominates the energy density of the universe are known to lead to 
adiabatic density perturbations. We show that generically the annihilation 
cross section of the same particle also receives fluctuations, which leads to 
entropy perturbations at freeze-out.  If the particle comes to dominate the
energy density of the universe and subsequently decays, this leads to an 
additional source of adiabatic density perturbations.  On the other hand, 
non-adiabatic density perturbations result when the particle does not decay 
but contributes to the observed dark matter.
\end{abstract}

\maketitle

\section{Introduction}
\label{sec:introduction}

Measurements of the cosmic microwave background radiation \cite{cmb1}
have revealed a highly uniform energy density background with super-horizon 
perturbations on the order of one part in $10^5$.  In the standard 
inflationary paradigm \cite{ref,infref}, these density perturbations were 
created in the inflationary epoch when quantum fluctuations of the inflaton 
field expanded beyond the Hubble radius and were converted into density 
perturbations upon inflaton decay.  However, to obtain the observed level of 
density perturbations from this mechanism requires tight constraints on the 
inflaton potential \cite{cmb2}.

Recently, Dvali, Gruzinov, Zaldarriaga and independently Kofman (DGZK) 
proposed a new mechanism \cite{DGZ} for producing density perturbations.
A nice feature of their scenario is that the only requirements on the 
inflaton potential are to produce the required e-foldings of inflation and 
at a scale consistent with WMAP data.  The 
DGZK mechanism posits the existence of some heavy particle $S$ with a mass 
and decay rate that depend on the vacuum expectation value of 
some light field $\chi$.  Here $\chi$ is presumed to have acquired 
super-horizon fluctuations during the inflationary epoch; however $\chi$ 
never contributes significantly to the energy density of the 
universe\footnote{The scenario where $\chi$ contributes significantly toward 
the energy density is called the ``curvaton'' scenario and was first proposed 
in \cite{curv}.}.  Nevertheless, the fluctuations in $\chi$ persist and 
result in fluctuations in the mass and decay rate of $S$, so long as the 
$\chi$ mass $m_\chi$ is less than the Hubble rate $H$ at the time at which 
fluctuations are transferred to radiation.  In the DGZK mechanism the 
field $S$ comes to dominate the energy density of the universe and decays 
into radiation while $m_\chi < H$.  Fluctuations in the mass and decay rate 
of $S$ result in fluctuations in the duration of $S$ energy domination, which 
in turn lead to adiabatic density perturbations since the energy of a massive 
$S$ field redshifts more slowly than that of radiation.  

The DGZK mechanism has been studied extensively.  For example, the evolution
of the density perturbations that result from this mechanism has been 
studied in detail using gauge invariant formalisms in \cite{gaugeinv}.  These 
perturbations are shown to possess a highly scale invariant spectrum in 
\cite{scaleinv} and are shown to contain significant non-Gaussianities in 
\cite{nongauss}.  The original DGZK mechanism has also been extended to apply 
to preheating as studied in \cite{extensions}.  For 
discussions of the limitations of this mechanism see for example 
\cite{limitations}.    

In the original DGZK scenario \cite{DGZ} it is assumed that $S$ decouples 
while being relativistic.  In this paper we generalize this to apply to the 
case where $S$ freezes-out of equilibrium with a fluctuating annihilation rate 
$\langle\sigma v\rangle$.  We use the term ``freeze-out'' to refer 
specifically to the scenario where $S$ decouples from thermal equilibrium 
after it has become non-relativistic.  In this case the number density of $S$ 
at a temperature $T$ after freeze-out is
\bea
n_S &\simeq& \frac{T^3}{m_Sm_{\rm pl}\langle \sigma v\rangle } \,,
\label{nSfreeze-out}
\eea
where $m_{\rm pl}$ is the Planck mass and $m_S$ is the mass of $S$.  
Therefore we expect fluctuations in the mass and annihilation rate of $S$ 
during freeze-out to result in fluctuations in the number density of $S$.  If 
$S$ lives long enough to dominate the energy density of the universe and 
subsequently decays, these entropy perturbations are converted into adiabatic 
perturbations.  These add to the ones produced by the original DGZK mechanism 
and the quantum fluctuations of the inflaton.

This paper is organized as follows.  In Section \ref{sec:analytical} we 
describe the density perturbations produced by our generalized DGZK mechanism. 
Sections \ref{sec:Smodels} and \ref{sec:chimodels} contain explicit models 
for implementing our mechanism and  for producing the fluctuating masses and 
coupling constants, respectively.  Conclusions are given in Section 
\ref{sec:conclusions}.  In Appendix  \ref{sec:pertanalysis} an alternate 
analytical description is given which allows to track the evolution of the 
perturbations, while in Appendix \ref{sec:numerical} Boltzmann equations are 
derived and solved numerically to confirm the analytical arguments presented 
in other sections of this paper.

\section{Analytical determination of the perturbations}
\label{sec:analytical}

Our  generalized DGZK mechanism includes a heavy particle $S$ with mass 
$m_S$, decay rate $\Gamma$ and annihilation cross section 
$\langle\sigma v\rangle$, where $S$ decays to and interacts with radiation. 
We begin by identifying several key temperature scales. The temperature at 
which $S$ begins to thermalize with radiation is denoted as $T_{\rm therm}$.  
We assume for simplicity that  $S$ particles are produced only as they 
thermalize from radiation annihilation below $T=T_{\rm therm}$.  We also 
define: 
\begin{enumerate}
\item $T_{\rm f.o.}$: Temperature at which $S$ freezes-out of thermal 
equilibrium;
\item $T_{\rm dom}$: Temperature at which $S$ begins to dominate the energy 
density of the universe;
\item $T_{\rm dec}$: Temperature at which $S$ decays.
\end{enumerate}
Since the number density of $S$ particles falls off exponentially after $S$
becomes non-relativistic, $T_{\rm f.o.}$ is typically within an order of 
magnitude of $m_S$.  Therefore in this paper we always take 
$T_{\rm f.o.}\simeq m_S$.  In terms of $m_S$, $\Gamma$ and 
$\langle\sigma v \rangle$ we also find
\bea
T_{\rm dom} &\simeq& \frac{1}{m_{\rm pl} \langle\sigma v\rangle} \nn\\
T_{\rm dec} &\simeq& m_{\rm pl} \, \Gamma^{2/3}\langle\sigma v\rangle ^{1/3}\,,
\label{Tdefs}
\eea
where we have assumed $T_{\rm dec}<T_{\rm dom}$ in the last equation.  This 
condition is necessary for significant density perturbations to be produced 
by this mechanism.  In Eq.~(\ref{Tdefs}) the cross section is to be evaluated 
at the freeze-out temperature $T_{\rm f.o.}$.  Note that for $S$ particles to 
be produced in the first place we require $T_{\rm therm}>T_{\rm f.o.}$.

As described in \cite{DGZ}, the period of $S$ domination between 
$T_{\rm dom}$ and $T_{\rm dec}$ gives rise to an enhancement of the resulting
energy density compared to a scenario where the $S$ domination is absent.  
Comparing energy densities at common scale factor one finds that after $S$ 
decays
\bea
\rho &=& \left( \frac{\rho_{\rm dom}}{\rho_{\rm dec}} \right)^{1/3} \!\!\!\!
\rho_{\rm rad}\,,
\label{rhoDGZ}
\eea
where
\bea
\rho_{\rm dom} &\simeq& T_{\rm dom}^4\,, \qquad 
\rho_{\rm dec} \,\simeq\, \frac{T_{\rm dec}^3}{m_{\rm pl} 
\langle\sigma v\rangle} \,,
\label{rhodefs}
\eea
and $\rho_{\rm rad}$ is the energy density which would result without any 
period of matter domination.  As discussed in detail in Section IV, couplings 
to an additional field $\chi$ can give rise to fluctuations in $m_S$, 
$\Gamma$, and $\langle \sigma v\rangle $:
\bea
m_S &=&\overline{m}_S \left( 1 + \delta_m \right) \nn\\
\langle\sigma v\rangle &=& \langle\overline{\sigma v}\rangle 
\left( 1 + \delta_{\langle\sigma v\rangle} \right) \nn\\
\Gamma &=& \overline\Gamma\left( 1 + \delta_\Gamma\right) \,,
\label{pertdefs}
\end{eqnarray}
where the barred quantities refer to background values.  According to 
Eqs.~(\ref{Tdefs}-\ref{pertdefs}), these fluctuations give rise to 
fluctuations in $T_{\rm dom}$ and $T_{\rm dec}$ which result in energy 
density perturbations
\bea
\frac{\delta\rho}{\rho} &=& -\frac{2}{3}\,\delta_\Gamma 
-\frac{4}{3}\,\delta_{\langle\sigma v\rangle} \,.
\eea
Note that although $\delta\rho/\rho$ contains no explicit dependence on 
$\delta_m$, both $\delta_{\langle \sigma v\rangle }$ and $\delta_\Gamma$ are 
in general functions of $\delta_m$.

Comparing the energy density at a common scale factor corresponds to choosing 
a gauge where the perturbation in the scale factor vanishes, $\psi=0$. Thus 
the fluctuation in the energy density computed here can be directly related 
to the gauge invariant Bardeen parameter \cite{bardeen}
\bea
\zeta &=& -\psi+\frac{\delta\rho}{3( \rho + p)} \,.
\label{zeta}
\eea
Thus we find after $S$ decays
\bea
\zeta = -\frac{1}{6}\,\delta_\Gamma 
-\frac{1}{3}\,\delta_{\langle\sigma v\rangle} \,.
\label{pertresults}
\eea

We can obtain the same result in synchronous gauge, where different 
regions all have the same global time.  Since $\rho \sim 1/t^2$ in  both matter and radiation dominated universes, one finds that $\delta\rho=0$ on surfaces of constant time. Thus
the Bardeen parameter is
\bea 
\zeta &=& -\psi = \frac{\delta a}{a}\,.
\label{uniHubblezeta}
\eea
To obtain $\zeta$, we only need to determine 
$a(t,\Gamma,\langle\sigma v\rangle,m)$ and then compare two regions at fixed 
$t$, but different $\Gamma$, $\langle\sigma v\rangle$ and $m_S$. 
Assuming the $S$ particles freeze-out while non-relativistic and decay after 
dominating the energy density of the universe, this gives 
\bea
a(t) &=& \frac{a(t)}{a(t_{\rm dec})} \frac{a(t_{\rm dec})}{a(t_{\rm dom})}
\frac{a(t_{\rm dom})}{a(t_{\rm f.o.})}\frac{a(t_{\rm f.o.})}{a(t_{0})}a(t_0) 
\nn\\ 
&=& \left(\frac{t}{t_0}\right)^{1/2} 
\left(\frac{t_{\rm dec}}{t_{\rm dom}}\right)^{1/6} a(t_0) \,,
\eea 
where $t_{\rm dec}\simeq\Gamma^{-1}$ is the time when $S$ decays, 
$t_{\rm dom}\simeq m^3_{\rm pl} \langle\sigma v\rangle ^2$ is the time at 
which it dominates the energy density of the universe, and  $t_{\rm f.o.}$ 
is the time at which it freezes-out.  Substituting gives 
\bea 
a(t) &=& \left(\frac{t}{t_0}\right)^{1/2} m^{-1/2}_{\rm pl} 
\Gamma^{-1/6} \langle\sigma v\rangle^{-1/3} a(t_0) \,.
\eea 
Using this result and Eq.~(\ref{uniHubblezeta}) we again obtain 
Eq.~(\ref{pertresults}).

The above discussion is approximate and requires that $S$ completely 
dominates the energy density of the universe.  Obtaining the perturbations 
when $S$ does not dominate requires that we include the matter contribution 
to the scale factor or energy density during radiation domination. This is 
done in Appendix \ref{sec:pertanalysis} using a different formalism. In 
Appendix \ref{sec:numerical} we confirm these analytic results using a 
numerical calculation of the density perturbations using Boltzmann equations.

\section{Explicit Models for Coupling $S$ to Radiation}
\label{sec:Smodels}

It is important to verify that models exist which exhibit the features 
discussed in the previous section. We present two models in which the 
annihilation cross section is determined by renormalizable and 
non-renormalizable operators, respectively. 

The first model is given by the Lagrangian
\bea
{\cal L} &=& \sqrt{-g} \left[ \frac{(\partial_\mu S)^2}{2} 
+\frac{(\partial_\mu X)^2}{2} -\frac{m_S^2}{2} S^2 -\frac{m_X^2}{2} X^2 
\right. \nn\\
& & \hspace{1.5cm} \left.  
- \frac{g\,m_S}{2}\,S\,X^2 -\frac{\lambda}{4}\,S^2 X^2 \right] \,.
\label{Smodel1}
\eea
We assume that $X$ is in thermal equilibrium with the remaining radiation 
and that $S$ particles are only produced through their coupling to $X$.  
The interaction terms in the above model yield an $S$ decay rate and 
cross section
\bea
\Gamma &\sim& g^2m_S \,,\qquad
\langle \sigma v\rangle  \,\sim\, \frac{\lambda^2}{M^2} \,,
\label{gammasigma}
\eea
where
\bea
M &\simeq& \left\{ 
\begin{array}{ll}
T & \,{\rm when}\,\,\, T \,>\, m_S \\
m_S & \,{\rm when}\,\,\, T \,<\, m_S
\end{array}
\right. \,.
\label{Mdef}
\eea
Note that we neglect the ${\cal O}(g^4)$ contribution to the cross section.  
This is justified given the limits on the coupling constants derived below. 

The requirement that $T_{\rm therm} > m_S$ and that $S$ remains in thermal 
equilibrium down to $T \simeq m_S$ gives the condition on the coupling 
$\lambda$
\bea
\lambda &>& \sqrt{\frac{m_S}{m_{\rm pl}}} \,.
\label{lambdaconstraint}
\eea
On the other hand the condition $T_{\rm dec} < T_{\rm dom}$ implies
\bea
g^2 \lambda^4 &<& \frac{m_S^3}{m_{\rm pl}^3}\,.
\label{glambdaconstraint}
\eea
Thus a necessary (but not sufficient) condition on $g$ to satisfy both 
Eq.~(\ref{lambdaconstraint}) and Eq.~(\ref{glambdaconstraint}) is
\bea
g &<&  \sqrt{\frac{m_S}{m_{\rm pl}}}\,.
\eea
Finally, we require that the period of $S$ domination does not disrupt 
big bang nucleosynthesis (BBN).  Thus the decay of $S$ must reheat the 
universe to a temperature $T_{\rm rh}>T_{\rm BBN}$, where 
$T_{\rm rh}\simeq \sqrt{\Gamma m_{\rm pl}}$.  This gives
\bea
g^2 &>& \frac{T_{\rm BBN}^2}{m_S m_{\rm pl}} \,.
\eea   
Using $T_{\rm BBN}\simeq 10^{-21}m_{\rm pl}$, the above relations provide 
the constraint $m_S\gtrsim 10^{-21}m_{\rm pl}$.  Given any $m_S$ satisfying 
this constraint, limits on $\lambda$ and $g$ are calculated using 
Eq.~(\ref{lambdaconstraint}) and Eq.~(\ref{glambdaconstraint}).  

Note that in this model the $S$ particles are produced at 
$T = T_{\rm therm}$ and remain in thermal equilibrium with the radiation 
until they freeze-out at $T \simeq m_S$. This is different from the 
assumption made in \cite{DGZ}, where $S$ starts in thermal equilibrium and 
decouples while still relativistic. In order to achieve this scenario, the 
coupling of $S$ to radiation has to proceed via a higher dimensional 
operator, or in other words via the propagation of an intermediate particle 
with mass much greater than $m_S$. 

This brings us to our second model. Consider a heavy fermion $\psi_S$ and
a light fermion $\psi_X$, coupled via an additional heavy scalar $\phi_H$ 
with mass $m_H$,
\begin{eqnarray}
{\cal L}_{\rm int} = g_S\,\bar{\psi}_S\psi_S\,\phi_H 
+ g_X\,\bar{\psi}_X\psi_X \, \phi_H\,.
\end{eqnarray}
We also assume that the fermion $\psi_S$ decays to radiation with rate 
$\Gamma$. The annihilation cross section is given by
\begin{eqnarray}
\langle \sigma v \rangle \sim  \frac{g_S^2 g_X^2}{m_H^4} M^2\,,
\end{eqnarray}
where $M$ is defined in Eq.~(\ref{Mdef}). 
In this case, thermalization occurs for temperatures bounded by
\begin{eqnarray}
m_S \left( \frac{m_H^4}{m_S^3 m_{\rm pl}} \frac{1}{g_S^2 g_X^2} \right)^{1/3} 
< T < g_S^2 g_X^2 m_{\rm pl}\,.
\end{eqnarray}
The conditions that $S$ is in in thermal equilibrium when it reaches 
$T \simeq m_S$ gives the condition
\begin{eqnarray}
g_S^2 g_X^2 > \frac{m_H^4}{m_S^3 m_{\rm pl}}\,.
\label{gSgXlimit}
\end{eqnarray}
Note that one still needs to have a decay rate that is small enough such that 
$\psi_S$ decays after it dominates the universe. The point of this second 
example is to show that in non-renormalizable models the heavy species can 
either decouple while non-relativistic or while relativistic, depending on 
whether Eq.~(\ref{gSgXlimit}) is satisfied or not.

\section{Models for producing the fluctuations}
\label{sec:chimodels}

The density perturbations in the DGZK mechanism and our generalization
originate in fluctuations in a light scalar field $\chi$.  In this section we 
write down explicit models for couplings between $S$ and $\chi$.  The reason 
for doing this is that these interactions can give rise to back reactions 
which can constrain the magnitude of the produced density perturbations.  
Similar results hold for couplings between $\psi_S$ and $\chi$. 

We find it convenient to define $\delta \chi\equiv \chi - \langle\chi\rangle$.  
Note that this does not correspond to a perturbative expansion. The fluctuations in $\chi$ 
are created during the inflationary era with
$\delta\chi\sim H_{\rm inf}$.  
Then the leading order equation of motion for $\chi$ can be split into 
homogeneous and inhomogeneous parts, 
\bea
\langle \ddot{\chi}\rangle &=& -3H\langle\dot{\chi}\rangle-
\langle V' \rangle \,,\nn\\
\delta\ddot{\chi} &=& -3H\delta\dot{\chi}+4\dot{\phi}\langle\dot{\chi}\rangle
-\delta V'-2\phi \langle V' \rangle \,.
\label{chieom}
\eea  
Here $\delta V' \equiv V'-\langle V'\rangle$, where $V$ is the potential of $\chi$ 
and the prime denotes a derivative 
with respect to $\chi$.  Also, $\phi$ is the time perturbation in conformal 
Newtonian gauge.  The terms proportional to $\phi$ enter into the leading 
order equation of motion for $\delta\chi$ because their homogeneous 
coefficients do not.  

To simplify the analysis, we first consider the scenario where 
$\langle\chi\rangle$ is negligible.  From Eqs.~(\ref{chieom}) we see this
is the case when $\langle\chi\rangle<\delta\chi$.  Thus we require the 
equation of motion for $\langle\chi\rangle$ to be Hubble friction dominated 
for $\langle\chi\rangle<\delta\chi$.  This gives the condition
\bea
H^2\delta\chi &>& H^2\langle\chi\rangle \,\,>\,\, \langle V' \rangle \,.
\label{simpcond}
\eea
The fluctuations $\delta\chi$ persist so long as the equation of motion 
for $\delta\chi$ is Hubble friction dominated.  With 
$\langle\chi\rangle<\delta\chi$ this translates into the condition
\bea
H^2\delta\chi &>& \delta V' +2\phi\langle V'\rangle \,.
\label{chiweakmasscond}
\eea  
Note that we can combine our simplifying condition that $\langle\chi\rangle$
be negligible, Eq.~(\ref{simpcond}), with the condition that the fluctuations
in $\delta\chi$ be Hubble friction dominated, Eq.~(\ref{chiweakmasscond}).  
Adding these two equations and dropping factors of 2 
this gives the single condition 
\bea
H^2\delta\chi &>& V' \,.
\label{chimasscond}
\eea

We consider the constraints this condition imposes on models for
transferring $\chi$ fluctuations to the radiation.  We first consider  
the renormalizable interactions
\bea
{\cal L}_{\chi} &=& \sqrt{-g}\left[ -\frac{\alpha_S}{4} S^2\chi^2 
- \frac{\mu_S}{2} S^2\chi \right] \,,
\label{chifluctmodels1}
\eea 
and neglect any couplings between $\chi$ and $X$ as they are irrelevant to
our mechanism.  When $\chi$ fluctuates these interactions result in $S$ mass 
fluctuations of
\bea
\delta_m &=& \frac{\alpha_S\delta\chi^2}{4m_S^2} 
+ \frac{\mu_S\delta\chi}{2m_S^2} \nn\\
&\sim& \sqrt{\left(\frac{\alpha_S H_{\rm inf}^2}{m_S^2}\right)^2 \!\!
+ \left(\frac{\mu_S H_{\rm inf}}{m_S^2}\right)^2}\,,
\label{mfluct}
\eea
where in the second line we have estimated the size of the rms fluctuation at 
two widely separated co-moving points.  This mass fluctuation gives rise to 
fluctuations in the decay rate and the annihilation cross section of $S$ 
according to the mass dependence of Eqs.~(\ref{gammasigma}).  

As described above, for this fluctuation to persist and for 
$\langle\chi\rangle$ to remain negligible requires that 
$H^2\delta\chi >V'$.  Although we assume the self interaction of $\chi$ 
is always negligible, the interactions of ${\cal L}_\chi$ contribute to 
$V$ and provide the constraint 
\bea
H^2\delta\chi &>& \left( \frac{\alpha_S}{2}\delta\chi 
+ \frac{\mu_S}{2} \right) \langle S^2 \rangle \,,
\eea  
where $\langle S^2 \rangle$ is evaluated in the thermal bath.  This 
constraint is tightest at $T=m_S$ when 
$\langle S^2 \rangle/H^2\sim m_{\rm pl}^2/m_S^2$.  Thus we obtain the 
constraints
\bea
\alpha_S &<& \frac{m_S^2}{m_{\rm pl}^2} \,,\qquad
\mu_S \,\,<\,\, \frac{m_S^2}{m_{\rm pl}^2}H_{\rm inf} \,.
\label{alphamuconstraints}
\eea

The constraints of Eqs.~(\ref{alphamuconstraints}) provide the same upper 
bound to both terms in Eq.~(\ref{mfluct}).  Thus the back reactions of 
${\cal L}_\chi$ limit the level of density perturbations produced via 
this mechanism to
\bea
\zeta &\sim& \delta_m \,<\, \frac{H_{\rm inf}^2}{m_{\rm pl}^2} 
\,\lesssim\, 10^{-8}\,,
\label{zetaconstr1}
\eea
where the last limit on $H_{\rm inf}/m_{\rm pl}$ is measured by the WMAP 
collaboration \cite{cmb2}.  

The fluctuations resulting from the second interaction in ${\cal L}_\chi$
are linear in $\delta\chi$ and are therefore predominantly Gaussian in 
their distribution.  Since the observed level of Gaussian fluctuations 
sets $\zeta\sim 10^{-5}$, this interaction cannot provide a significant 
fraction of the observed density perturbations.  However, the fluctuations
resulting from the first term in ${\cal L}_\chi$ are quadratic in 
$\delta\chi$  and therefore non-Gaussian \cite{extensions}.  Recent analysis
\cite{cmb3} limits the amplitude of non-Gaussian perturbations to about 
$10^{-8}$.  Thus we see our model can provide non-Gaussian perturbations 
right at the limit of current observation.  A lower level of 
perturbations is obtained by reducing $\alpha_S$ or $\mu_S$.

As a variant on the above scenario, we next consider the non-renormalizable 
couplings
\bea 
{\cal L}'_{\chi} &=& \sqrt{-g} \left[ -\frac{\lambda}{4}\,
\frac{\chi^2}{M_1^2}\,S^2 X^2 -\frac{\lambda}{4}\,
\frac{\chi}{M_2} S^2X^2\right] \,.
\label{chifluctmodel2}
\eea 
When $\chi$ fluctuates these interactions 
result in fluctuations in $\langle \sigma v\rangle $
\bea 
\delta_{\langle \sigma v\rangle } &=& \frac{2\delta \chi^2}{M_1^2 } 
+ \frac{2\delta \chi}{M_2} 
\,\sim\, \sqrt{ \left(\frac{ H_{\rm inf}^2}{M_1^2}\right)^2 \!\!
+ \left(\frac{ H_{\rm inf}}{M_2}\right)^2} . \quad
\label{svfluct}
\eea 

As above, we require that Eq.~(\ref{chimasscond}) be satisfied.  For the 
interactions of ${\cal L}'_\chi$ this gives
\bea
H^2\delta\chi &>& \frac{\lambda}{4}\left( 
\frac{2\delta\chi}{M_1^2}+\frac{1}{M_2}
\right) \langle S^2X^2 \rangle \,.
\eea
As in the previous example, this constraint is tightest at $T=m_S$ when 
$\langle S^2X^2 \rangle/H^2\sim m_{\rm pl}^2$.  Therefore we find
\bea
\frac{1}{M_1^2} &<& \frac{1}{\lambda}\frac{1}{m_{\rm pl}^2} \,,\qquad
\frac{1}{M_2} \,\,<\,\, \frac{1}{\lambda}\frac{H_{\rm inf}}{m_{\rm pl}^2} \,.
\label{Mconstraints}
\eea

Analogous to the previous example, the constraints of 
Eqs.~(\ref{Mconstraints}) provide the same upper bound to both terms in 
Eq.~(\ref{svfluct}).  Thus the back reactions of ${\cal L}'_\chi$ limit the 
level of density perturbations produced via this mechanism to
\bea
\zeta &\sim& \delta_{\langle \sigma v\rangle } 
\,<\, \frac{1}{\lambda}\frac{H_{\rm inf}^2}{m_{\rm pl}^2} \,. 
\label{zetaconstr2}
\eea
This bound is significantly weaker than the bound of Eq.~(\ref{zetaconstr1})
obtained via a fluctuating $S$ mass.  For example, the fluctuations resulting
from ${\cal L}'_\chi$ could form the dominant contribution to the observed
density perturbations if $\lambda$ is sufficiently small.  In addition, for
a given $\zeta$ decreasing $\lambda$ allows for a lower scale of inflation.
Constraints on the smallness of $\lambda$ are discussed in Section 
\ref{sec:Smodels}.  Of course, a lower level of Gaussian (non-Gaussian) 
perturbations is obtained by increasing $M_2$ ($M_1$).  

Above we have taken $\langle \chi \rangle$ to be negligible, which 
corresponds to taking $\langle \chi\rangle<\delta\chi$.  Although this simplifies the presentation, 
it unnecessarily strengthens the
constraints on $\mu_S$ and $M_2$.  We know from Eq.~(\ref{simpcond}) that
\begin{eqnarray}
\langle \chi \rangle > \frac{\langle V' \rangle}{H^2}\,,
\end{eqnarray}
therefore keeping $\langle \chi \rangle$ small implies constraints on the potential $V$. 
Referring to the second of 
Eqs.~(\ref{chieom}), we see that for arbitrary $\langle\chi\rangle$ the 
requirement that $\delta\chi$ remains Hubble friction dominated gives
\bea
H^2\delta\chi \,>\, \dot{\phi}\langle\dot{\chi}\rangle \,,\quad
H^2\delta\chi \,>\, \delta V' \,,\quad
H^2\delta\chi \,>\, \phi \langle V' \rangle \,.\,\,
\label{chiweakmasscond2}
\eea
The first condition provides the constraint 
$\langle\chi\rangle < \delta\chi/\phi$, with the evolution of $\phi\sim\zeta$
described in Appendix \ref{sec:pertanalysis}.  It is sufficient to take
$\phi\sim 10^{-5}$, which also ensures that the homogeneous correction that 
$\langle\chi\rangle$ provides to $m_S$ does not change $m_S$ by more than 
order unity\footnote{In Appendix \ref{sec:pertanalysis} we find that after
freeze-out $\phi$ evolves as $\phi\sim (\rho_S/\rho)\,\zeta_{\rm f}$, where 
$\zeta_{\rm f}\sim 10^{-5}$ is the final curvature perturbation.  Thus if 
we consider the scenario where $\chi$ fluctuations are transferred at 
freeze-out and $\chi$ subsequently decays, we may take $\langle\chi\rangle$
to be constrained by $\phi^{-1}$ at freeze-out, which considerable weakens
the bounds in Eqs.~(\ref{newzeta}).  However, in this case 
$\langle\chi\rangle$ provides a homogeneous adjustment to $m_S$ which may 
be much larger than $m_S$.  This effect could then significantly alter the 
constraints calculated in Section \ref{sec:Smodels}.}.  
Through an analysis analogous to that above, we find the conditions of 
Eqs.~(\ref{chiweakmasscond2}) constrain the level of {\em Gaussian} 
fluctuations for the respective interactions of ${\cal L}_\chi$ and 
${\cal L}'_\chi$ to 
\bea
\zeta_{\rm g} \,<\, \frac{\langle\chi\rangle}{\delta\chi}
\frac{H_{\rm inf}^2}{m_{\rm pl}^2}\,,\quad
\zeta_{\rm g} \,<\, \frac{1}{\lambda}\frac{\langle\chi\rangle}{\delta\chi}
\frac{H_{\rm inf}^2}{m_{\rm pl}^2} \,;\quad
\frac{\langle\chi\rangle}{\delta\chi} \,<\, \frac{1}{\phi} \,.
\label{newzeta}
\eea
The additional factor of $\langle\chi\rangle/\delta\chi$ significantly 
weakens both bounds on Gaussian perturbations.  This allows for greater 
freedom in choosing $\mu_S$, $M_2$, $\lambda$, and/or $H_{\rm inf}$.  

Non-Gaussian perturbations originate from the couplings quadratic in $\chi$. Taking these into account the fluctuation resulting from 
${\cal L}_\chi$ becomes
\bea
\zeta \sim \delta_m = \frac{\alpha_S\delta\chi^2}{4m_S^2} 
+ \frac{\alpha_S\langle\chi\rangle\delta\chi}{2m_S^2}
+ \frac{\mu_S\delta\chi}{2m_S^2} \,.
\eea 
Note that the quadratic term $\sim \alpha_S$ also gives rise to a Gaussian contribution to $\zeta$. Thus, the non-Gaussian perturbations obey the relation
\begin{eqnarray}
\zeta_{\rm ng} \sim \frac{\alpha_S \, \delta \chi}{\mu_S + \alpha_S \langle \chi \rangle} \zeta_{\rm g}\,.
\label{zetangrelation}
\end{eqnarray}
Note that taking $\alpha_S \to 0$ the non-Gaussian fluctuations can be made arbitrarily small. However, even if the $\alpha_S$ term in the potential dominates, the non-Gaussian fluctuations are always limited by
\begin{eqnarray}
\zeta_{\rm ng} < \frac{H_{\rm inf}^2}{m_{\rm pl}^2}\,,
\end{eqnarray}
as can be seen by combining Eqs.~(\ref{newzeta}) and~(\ref{zetangrelation}). 
As mentioned before, WMAP sets the limit $H_{\rm inf}^2/m_{\rm pl}^2 \lesssim 10^{-8}$~\cite{cmb2}, thus the non-Gaussian fluctuations are at or below the current limits from WMAP~\cite{cmb3}. 	In addition, the observed Gaussian fluctuations can be produced by choosing $\delta \chi / \langle \chi \rangle$ appropriately. The second model with Lagrangian given in Eq.~(\ref{chifluctmodel2}) is less constrained since the factor of $\lambda$ weakens the constraint on $\delta \chi / \langle \chi \rangle$.

\section{Conclusions}
\label{sec:conclusions}

In \cite{DGZ} it was shown that fluctuations in the mass and the decay rate 
of a heavy particle $S$, which at some point dominates the energy density 
of the universe, lead to adiabatic density perturbations. In this scenario 
it was assumed that the heavy particle decouples from radiation while it is 
still relativistic. 

In this work we have shown that if the heavy particle remains in thermal 
equilibrium until it becomes non-relativistic, fluctuations in the 
annihilation cross section of this particle with radiation lead to 
additional sources of perturbations.  We have presented two simple toy 
models illustrating this effect. These additional fluctuations are generic, 
unless the annihilation cross section is mediated by an additional particle 
with mass exceeding $m_S$. If the $S$ particle is stable, for example if 
$S$ is dark matter, then the resulting perturbations are non-adiabatic.

A simple analytical calculation determines the size of the density 
perturbations from fluctuations in the mass, decay rate and annihilation 
cross section.  The fluctuations due to variations in the annihilation cross 
section are shown to be of similar size as the ones generated from the 
original DGZK mechanism.  These results are checked numerically using 
Boltzmann equations in conformal Newtonian gauge in 
Appendix~\ref{sec:numerical}.

\begin{acknowledgments}
We would like to thank Mark Wise for collaboration at an early stage of 
this work.  This work was supported by the Department of Energy under the 
contract DE-FG03-92ER40701.
\end{acknowledgments}

\appendix
\section{Evolution of Density Perturbations}
\label{sec:pertanalysis}

In this appendix we determine the evolution of density perturbations
generated by a fluctuating cross section or mass during freeze-out. 
Unlike other analytic derivations given elsewhere in this paper, the one 
provided here allows us to easily follow the growth of the non-adiabatic 
perturbation during the radiation dominated era after freeze-out. We work 
in conformal Newtonian gauge with negligible anisotropic stress and use the 
line element
\begin{equation}
ds^2 =a^2 \left[ -(1 +2 \phi) d\eta^2 + (1 -2 \psi) \delta _{ij} d x^i d x^j
\right]\,,
\label{metric}
\end{equation}
with $\psi = \phi$. We track the evolution of perturbations using the gauge 
invariant entropy perturbation ${\cal S}$ and curvature perturbation $\zeta$. 

We will assume that the scattering and annihilation interactions between $S$ 
and the radiation conserve total particle number.  This allows us to obtain 
a first integral of the Boltzmann equations. Conservation of total particle 
number in a fixed co-moving volume implies for the entropy perturbation
\begin{equation}
{\cal S} \equiv \delta_S - \delta _R = -\left( \delta_R - 3 \phi\right)
\left( 1+ \frac{n_R}{n_S}\right) + \frac{\lambda_0}{n_S a^3}\,,
\label{constraintS}
\end{equation}
where $\delta_S\equiv \delta n_S/n_S$ and $\delta_R\equiv \delta n_R/n_R$ are 
the perturbations in the number densities of $S$ and radiation, respectively. 
Here $\lambda_0$ is an integration constant which vanishes  in the absence of 
initial adiabatic perturbations. 

Eq.~(\ref{constraintS}) has a few salient features that we now discuss. First 
note that it admits an adiabatic solution ${\cal S}=0$ whenever  both 
$n_R/n_S$ and $n_S a^3$ are constant. This solution is the familiar
$\delta_S=\delta_R=3 \phi +{\rm constant}$, with the constant fixed by 
$\lambda_0$.  During freeze-out, however, the conditions described above 
are not satisfied, and entropy perturbations ${\cal S}^{\rm f.o.}$ are 
generated at that time. Using Eq.~(\ref{nSfreeze-out}) we find
\begin{equation}
\label{Sfo}
{\cal S^{\rm f.o.}} \simeq -\delta _{\langle\sigma v\rangle} -\delta _m ~.
\end{equation}
After 
freeze-out the heavy particle no longer interacts with the radiation and
therefore both $\delta_S$ and $\delta_R$ obey the perturbed Einstein field 
equation $\dot{\delta}_{S,R}= 3 \dot{\phi} $. This is trivially integrated 
to give
\begin{equation}
\delta _{S,R} = 3 \phi + \delta_{S,R} ^{\rm f.o.} -3 \phi^{\rm f.o.} \,.
\label{deltaSdec}
\end{equation}
Thus, after freeze-out the entropy perturbation remains constant 
${\cal S} = {\cal S}^{\rm f.o.}$. 

To derive the evolution of the curvature perturbation $\zeta$, we use 
Eq.~(\ref{zeta}) together with $\delta \rho_S = \rho_S(\delta _S + \delta_m)$
and $\delta \rho_R= (4/3) \rho_R \delta _R$ to find after several lines of 
algebra 
\begin{equation}
\zeta \simeq 
\frac{\rho_S( {\cal S}^{\rm f.o.} + \delta_m )}{3\rho _S + 4\rho_R}
+ \frac{1}{3} \left( \delta_R^{\rm f.o} - 3 \phi^{\rm f.o.} \right) \,.
\label{zetaapp}
\end{equation}
Using Eq.~(\ref{constraintS}), the second term on the right is suppressed by
$(n_S/n_R)(4\rho_R/\rho_S+3)$ compared to the first term and is subsequently
neglected. We see that in the radiation dominated era the curvature 
perturbations produced during freeze-out are suppressed relative to the 
entropy perturbations.

Once the the heavy particle starts to dominate the energy density of the 
universe, we find
\begin{equation}
\zeta \simeq -\frac{1}{3} \delta_{\langle \sigma  v \rangle} \,,
\label{zetafinal}
\end{equation}
in agreement with the results obtained in the main body of this paper.
If $S$ is the dark matter, such a large ratio of entropy to curvature 
perturbations is disfavored by data, which requires 
$|{\cal S}/\zeta|\lesssim 1/3$~\cite{cmb2}.  If instead $S$ decays, the 
entropy perturbations in $S$ are transferred to radiation.  The resulting 
curvature perturbation is then
\begin{equation}
\zeta \simeq -\frac{1}{3} \delta_{\langle \sigma  v \rangle} -\frac{1}{6}
\delta_{\Gamma} ~.
\end{equation}
Here we have finally included the effect of inhomogeneous $S$ decay,
which are formally calculated in~\cite{DGZ,gaugeinv}.

\section{Numerical Results}
\label{sec:numerical}

We now solve for the evolution of density perturbations during freeze-out 
using Boltzmann equations (see for example \cite{ref,MB}).  
As above, we work in conformal Newtonian gauge and consider
only super-horizon perturbations, neglecting all spatial gradients next to
conformal time derivatives.  Since we assume that $S$ is non-relativistic, 
the distribution function for $S$ is
\bea
f_S = e^{\mu/T} f_{\rm eq}\,,
\eea
where $f_{\rm eq}$ is the Maxwell-Boltzmann equilibrium distribution function. 
Then integrating the background Boltzmann equations over all of phase space
gives 
\bea
\dot{n}_S+3{\cal H} n_S &=& a\langle \overline{\sigma v}\rangle 
\left( n_{\rm eq}^2-n_S^2 \right) -a\overline{\Gamma}n_S \nn\\
\dot{n}_R+3{\cal H} n_R &=& -\,a\langle \overline{\sigma v}\rangle 
\left( n_{\rm eq}^2-n_S^2 \right) +a\overline{\Gamma}n_S \,.
\label{Bzero}
\eea
where dots denote derivatives with respect to conformal time, 
${\cal H}=\dot{a}/a$ and $n_{\rm eq}$ is the equilibrium number density of 
$S$ particles.  To describe the evolution of the scale factor we use 
Einstein's field equation,
\bea
{\cal H}^2 &=& \frac{a^2}{3m_{\rm pl}^2}
\left( \rho_R+\rho_S \right) .
\label{Hzero}
\eea

\begin{figure}[t!]
\includegraphics[width=7.75cm]{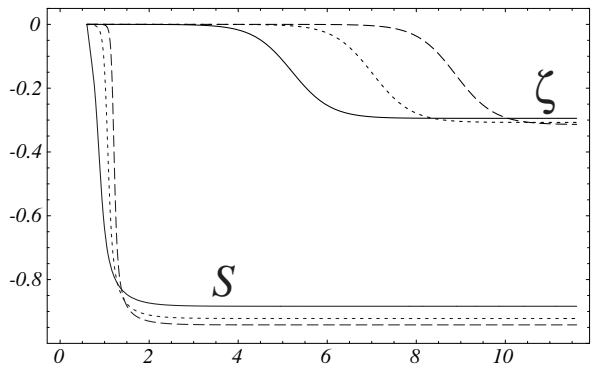}
\caption{Evolution of ${\cal S}$ and $\zeta$ in units of 
$\delta_{\langle \sigma v \rangle}$ as a function of $\log (m_S/T)$.  
The solid, dotted and dashed curves correspond to 
$\langle \sigma v \rangle m_{S}m_{\rm pl} = 10^{5}$, $10^{7}$ and $10^{9}$, 
respectively.  The large values of $\langle \sigma v \rangle m_{S}m_{\rm pl}$ 
are necessary to ensure that the heavy particle is in thermal equilibrium 
at $T_0 = m_S/4$.}
\label{fig1}
\end{figure}

To derive the subleading order Boltzmann equations we allow for fluctuations
in $m_S$, $\langle\sigma v\rangle$ and $\Gamma$ as defined in 
Eqs.~(\ref{pertdefs}).  Note that the resulting fluctuations in the number
density of the radiation $\delta_R$ correspond to temperature fluctuations 
$\delta_T=\delta_R/3$.  Therefore the equilibrium number density of $S$ 
particles acquires fluctuations 
\bea
\delta_{\rm eq} = \frac{3}{2} \left(\delta_m + \delta_T \right) + 
\exp\left[-\frac{m_S}{T}(\delta_m- \delta_T)\right] - 1 \,.
\eea
When the subleading Boltzmann equations are integrated over all phase space
we obtain
\bea
\dot{\delta}_S-3\dot{\phi} &=& a\langle\overline{\sigma v}\rangle 
\frac{n_{\rm eq}^2}{n_S}
\left( \phi+\delta_{\langle\sigma v\rangle }
+2\delta_{\rm eq}-\delta_S \right) \nn\\ 
& & -\,a\langle \overline{\sigma v}\rangle  
n_S\left( \phi+\delta_{\langle \sigma v\rangle }+\delta_S \right) 
-a\overline{\Gamma}\left( \phi+\delta_\Gamma\right) \nn\\
\dot{\delta}_R-3\dot{\phi} &=& -\,a\langle \overline{\sigma v}\rangle 
\frac{n_{\rm eq}^2}{n_R}
\left( \phi+\delta_{\langle \sigma v\rangle }
+2\delta_{\rm eq}-\delta_R \right) \nn\\
& & +\,a\langle \overline{\sigma v}\rangle \frac{n_S^2}{n_R}
\left( \phi+\delta_{\langle \sigma v\rangle }+2\delta_S-\delta_R \right) \nn\\
& & +\,a\overline{\Gamma}\frac{n_S}{n_R}
\left( \phi+\delta_\Gamma+\delta_S-\delta_R \right) \,.
\label{Bfirst}
\eea
Note that in deriving the Boltzmann equations we assumed 
$\delta_{\rm eq} \ll 1$. This ceases to be valid once 
$T \lesssim m_S (\delta_m - \delta_T)$. In this case, however, the factor of 
$n_{\rm eq}^2$ in front of the terms containing $\delta_{\rm eq}$ is 
exponentially suppressed compared to the remaining terms. 

The independent perturbations in Eqs.~(\ref{Bfirst}) are $\delta_S$, 
$\delta_R$, and $\phi$.  To describe the evolution of $\phi$ we use the 
first order perturbation to Einstein's field equation,
\bea
{\cal H}\dot{\phi}+{\cal H}^2\phi &=& -\frac{a^2}{6m_{\rm pl}^2}
\left( \delta\rho_R+\delta\rho_S \right) \,.
\label{Hfirst}
\eea

We solve the above system of equations numerically.  The effects of 
a fluctuating decay rate $\Gamma$ are well-studied \cite{DGZ,gaugeinv}
so we set $\Gamma=0$ to simplify our results.  For concreteness we also 
assume $S$ interacts with one out of one hundred radiative degrees of freedom 
and we begin integration at $T_0=m_S/4$ with $n_S(T_0)=n_{\rm eq}(T_0)$.  The 
results for the gauge invariant quantities ${\cal S}$ and $\zeta$ are shown in 
Fig.~\ref{fig1} for several values of 
$\langle \sigma v \rangle \, m_S\, m_{\rm pl}$. 

The curvature perturbation $\zeta$ is negligible compared to 
$\delta_{\langle \sigma v \rangle}$ until the heavy particle contributes 
significantly to the energy density. It asymptotes to a value 
$\zeta \simeq -0.3$. The entropy perturbation ${\cal S}$ grows during 
freeze-out and soon thereafter reaches a constant value of 
${\cal S} \simeq -0.9$. These results and the other features of 
Fig.~\ref{fig1} are in good agreement with the analytical results given 
in  Eqs.~(\ref{Sfo}) and~(\ref{zetafinal}).

\end{document}